\documentclass[preprint]{elsarticle}

\usepackage{geometry}
\usepackage{amsmath}
\usepackage{amsfonts}
\usepackage{booktabs}
\usepackage{pdfpages}
\usepackage{algorithm}
\usepackage{algpseudocode}
\usepackage{graphicx}
\usepackage{hyperref}
\usepackage{bm}

\usepackage{hyperref}

\journal{Journal of Intelligent Manufacturing}

\begin{document}

\begin{frontmatter}

\title{Intelligent Algorithms For Signature Diagnostics Of Three-Phase Motors}

\author[aff1]{Stepan Svirin\fnref{fn1}}
\author[aff1]{Artem Ryzhikov\fnref{fn2}}
\author[aff1]{Saraa Ali\fnref{fn3}}
\author[aff1]{Denis Derkach\fnref{fn4}}

\address[aff1]{National Research University Higher School of Economics, Moscow, Russia}

\fntext[fn1]{\texttt{stequoy@gmail.com}; ORCID: 0009-0008-9339-3515}
\fntext[fn2]{\texttt{aryzhikov@hse.ru}; ORCID: 0000-0002-3543-0313}
\fntext[fn3]{\texttt{thraaali@hse.ru}; ORCID: 0000-0002-1154-1262}
\fntext[fn4]{\texttt{dderkach@hse.ru}; ORCID: 0000-0001-5871-0628}

\begin{abstract}
The application of machine learning (ML) algorithms in the intelligent diagnosis of three-phase engines has the potential to significantly enhance diagnostic performance and accuracy. Traditional methods largely rely on signature analysis, which, despite being a standard practice, can benefit from the integration of advanced ML techniques. In our study, we innovate by combining state of the art algorithms with a novel unsupervised anomaly generation methodology that takes into account physics model of the engine. This hybrid approach leverages the strengths of both supervised ML and unsupervised signature analysis, achieving superior diagnostic accuracy and reliability along with a wide industrial application. Our experimental results demonstrate that this method significantly outperforms existing ML and non-ML state-of-the-art approaches while retaining the practical advantages of an unsupervised methodology. The findings highlight the potential of our approach to significantly contribute to the field of engine diagnostics, offering a robust and efficient solution for real-world applications.
\end{abstract}

\begin{keyword}
Induction Motor Fault Detection \sep Signature Analysis \sep Machine Learning \sep Generative Neural Networks
\end{keyword}

\end{frontmatter}

\section*{Acknowledgements}
The publication was supported by the grant for research centers in the field of AI provided by the Analytical Center for the Government of the Russian Federation (ACRF) in accordance with the agreement on the provision of subsidies (identifier of the agreement 000000D730321P5Q0002) and the agreement with HSE University No. 70-2021-00139. The funders had no role in study design, data collection and analysis, decision to publish, or preparation of the manuscript.

This work is supported by HSE Basic Research Fund. The computation for this research was performed using the computational resources of HPC facilities at HSE University.

\clearpage
\section{Introduction}

The reliability and efficiency of three-phase engines is critical for numerous industrial applications, which makes their accurate and timely diagnosis essential. Traditional diagnostic methods for these engines predominantly rely on signature analysis, a technique that examines the engine's operational patterns to detect anomalies~\cite{book_motor}. While signature analysis has become a de-facto standard due to its effectiveness, it has some substantial limitations, and the growing complexity of modern engines and the vast amounts of data they generate require more advanced and precise diagnostic frameworks~\cite{KHANJANI2021108622}.

At the same time, machine learning (ML) and artificial intelligence (AI) have emerged as essential tools integrated into various aspects of modern life, from recommendation algorithms~\cite{review_RS_ML} to healthcare~\cite{s23094178} applications. The potential for advancement and innovation in these fields is immense. Despite this, the application of ML in industrial settings remains underexplored, primarily due to the scarcity of publicly available labeled datasets, especially with malfunctioning engines~\. This lack of data poses significant challenges when transitioning ML solutions from experimental phases to full-scale production, especially given the complexities and variability of real-world conditions \cite{peres2020industrial}.

One critical area where ML can significantly impact industrial applications is in the diagnosis and maintenance of three-phase engines. These engines, also known as three-phase induction motors, are a cornerstone of industrial operations because of their robustness, efficiency, and reliability \cite{IMARC2023}. They are prevalent in numerous applications, including pumps, compressors, conveyors, and fans, making their accurate and timely diagnosis crucial to ensuring uninterrupted industrial productivity and safety \cite{Siddiqui2014}. While the use of ML-based fault detection in experimental setups provides effective and reliable solutions \cite{Kumar2021}, in the context of real-life implementation, this field remains poorly studied, highlighting a significant opportunity for innovation and improvement.

One of the main reasons for this is that most current ML approaches require labeled data, even for the binary classification of healthy and faulty time series, and the classification of specific fault types requires even more labeled data. This is problematic as companies are unwilling to damage expensive equipment to generate the necessary labeled datasets.

In this research, we investigate the detection of defects in three-phase induction motors using modern machine learning techniques, complemented by a novel signature-guided unsupervised anomaly generation methodology. Our goal is to review existing solutions for the fault detection problem based on time series data collected from engines, and to develop a robust approach that mitigates the need for extensive labeled datasets.

To address this challenge, we propose a novel approach that combines signature analysis with generative neural networks and ResNet-based \cite{he2015deepresiduallearningimage} convolutional neural networks to generate physically accurate synthetic anomalies. These synthetic anomalies are then used to train supervised machine learning models, enhancing their ability to detect and classify real-world anomalies. Our approach effectively integrates the strengths of both supervised and unsupervised methods, providing a practical solution for accurate and reliable fault detection in three-phase induction motors without dependence on large-labeled datasets. We have adapted the state-of-the-art unsupervised models AnomalyBERT \cite{jeong2023anomalybert} and VAE-LSTM \cite{VAE-LSTM} to serve as unsupervised baselines for induction motor fault detection. The experiments demonstrate a strong outperformance and potential of the proposed approach over the existing baselines.

The next section of the manuscript will formally define the problem at hand, introduce the existing solutions, and outline the methodology and limitations of the approaches explored and proposed. Next, the proposed algorithm is introduced. Finally, the manuscript concludes with the results and discussion, summarizing the findings and their implications.

\section{Background}
This section reviews state-of-the-art solutions related to the problem of fault detection in induction motors, focusing on signature analysis and machine learning-based approaches.

\subsection{3-phase engines}
Three-phase engines, also known as three-phase induction motors, are a fundamental component in various industrial applications due to their robustness, efficiency, and reliability. These motors operate on a three-phase power supply, which generates a rotating magnetic field within the stator. The stator is the stationary part of the motor and contains windings connected to the three-phase power supply. When the three-phase current flows through the stator windings, it induces a rotating magnetic field. The rotor, which is the spinning part of the motor, is placed within the stator. It typically consists of a laminated iron core with conductors, usually in the form of aluminum or copper bars, embedded within it. These conductors are arranged to form a closed loop known as a squirrel-cage rotor.
 
The operation of three-phase engines is based on the principles of electromagnetic induction. As the rotating magnetic field produced by the stator sweeps past the rotor, it induces an electromotive force (EMF) in the rotor conductors. According to Faraday's Law of Electromagnetic Induction, this induced EMF generates currents within the rotor. These currents interact with the magnetic field of the stator, producing a torque that causes the rotor to turn. The speed of the rotor is slightly less than the speed of the rotating magnetic field, a difference known as a slip, which is essential for torque production.

Three-phase induction motors are favored in industrial settings for several reasons. They offer high efficiency, providing a superior power-to-weight ratio compared to other motor types. Their simple and rugged construction makes them highly durable and capable of withstanding harsh operating conditions. Moreover, they require relatively low maintenance since they lack brushes and commutators found in other types of motors, reducing the wear and tear typically associated with these components.

The versatility and reliability of three-phase induction motors make them suitable for a wide range of applications. They are commonly used in machinery such as pumps, compressors, conveyors, and fans, where consistent and reliable operations are critical. Given their widespread use, the ability to accurately diagnose and maintain these motors is of paramount importance to ensure operational efficiency and minimize downtime in industrial processes. Traditional diagnostic methods, primarily based on signature analysis, have been effective but are increasingly challenged by the complexity and volume of data generated by modern engines. This requires the development of more advanced diagnostic tools, incorporating machine learning and artificial intelligence to enhance the precision and reliability of fault detection in three-phase induction motors.

\subsection{Engine Defect Types}

Three-phase induction motors, despite their robustness, can suffer from various defects that impact performance and lifespan \cite{Albrecht1986}. Understanding these defects is the key for effective diagnostics and maintenance. Common defect types include:

\subsubsection{Intercell Shortages}

Intercell shortages, which account for $42\%$ of overall faults \cite{Singh2003}, are the result of short circuits between the turns of the stator windings, typically due to insulation breakdown from thermal stress, electrical surges, or physical damage. This defect causes uneven current distribution, localized heating, increased losses, and reduced efficiency, potentially leading to catastrophic failure if unaddressed.

\subsubsection{Rotor Cell Defect}

Rotor cell defects involve faults in the rotor bars or end rings, caused by manufacturing flaws, thermal stress, or mechanical fatigue, account for approximately $10\%$ of all faults encountered in induction motors \cite{Nandi2005}. These defects, such as broken bars or cracked rings, disrupt the motor's electromagnetic balance, leading to increased vibration, reduced torque, and uneven rotor heating, accelerating wear on other components.

\subsubsection{Bearing Defect}

Bearing defects, which arise from inadequate lubrication, contamination, misalignment, or material fatigue \cite{Harris2006}, account for almost half of all induction motor failures \cite{report}. Symptoms include increased noise, vibration, and friction, leading to overheating and potential bearing seizure. Bearing failures can cause rotor misalignment and stator-rotor contact, necessitating costly repairs or replacements.

\subsubsection{Eccentricity of the Air Gap}

Eccentricity of the air gap refers to uneven spacing between the stator and rotor, caused by manufacturing inaccuracies, bearing wear, or rotor deformation \cite{Cameron1986}. It can be static (fixed) or dynamic (variable). This defect leads to unbalanced magnetic pull, increased vibration, noise, and uneven wear, resulting in performance degradation and potential failure over time.

\subsubsection{Unknown Mechanical Defects}

Unknown mechanical defects encompass issues like misalignments, loose components, structural weaknesses, and other unidentified problems. These defects often cause abnormal vibrations, noises, or irregular behavior, requiring comprehensive analysis and advanced diagnostics to accurately identify and rectify.

Each of these defect types presents unique challenges for the diagnosis and maintenance of three-phase induction motors. Developing advanced diagnostic methods, such as those incorporating machine learning and AI, can significantly enhance the detection and classification of these defects, leading to improved motor reliability and longevity.

\subsection{Existing diagnosis approaches. Signature Analysis}

Signature analysis is a theoretical approach that is often utilized in the task of induction motor fault detection. It involves examining the unique "signatures"\footnote{Patterns or sets of data that represent normal operational conditions.} and compares them with current operational data. These signatures can be extracted from different sources like vibration \cite{Raj2013}, pressure or acoustic data. While methods based on this data perform pretty well \cite{AlShorman2021}, current signature analysis (CSA) has some advantages over vibration or acoustic methods, as it is the most cost-effective and can be used in situations when direct physical contact with the engine is impossible \cite{Corne2015}. This efficiency stems from the ability to monitor the data by attaching current clamps to power wires, making the method non-invasive \cite{sunal2022review}. In CSA, the electrical current flowing through the stator is monitored.

This method works on the hypothesis that different types of fault in the motor generate a specific current signature, the harmonics of which can be observed in the stator current spectrum \cite{moiz2019health}. These signatures could be specific anomalies in the time series current data transformed into spectrum of a signal. For example, unexpected spikes indicating a fault. Unlike methods that may rely on a comparison with normal motor behaviour, the signature method predefines what anomaly signatures to look for based on known motor behaviors and known physical parameters. 

\begin{figure}
    \centering
    \includegraphics[width=0.7\linewidth]{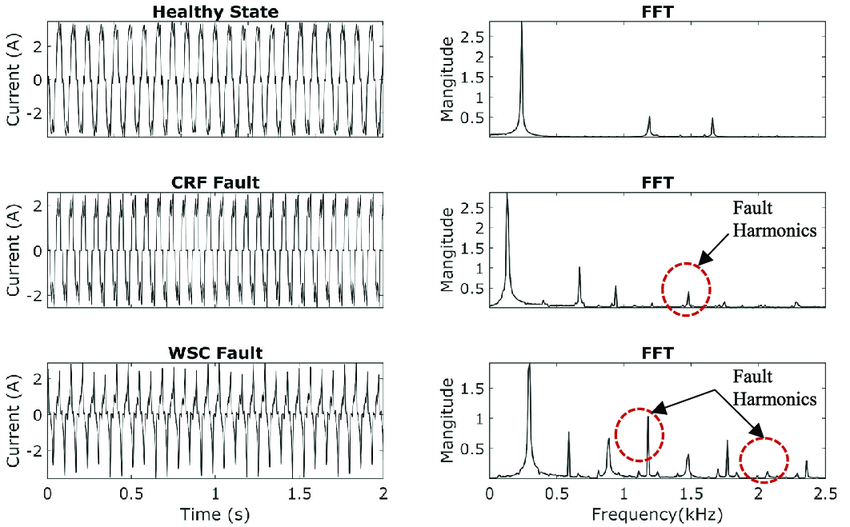}
    \caption{Signature method illustration~\cite{signature}. On the left side, the engine current over time is shown. On the right side, the corresponding Fourier spectrum of the current is displayed. The first row shows a healthy engine's current and spectrum, which has only a single peak at the operating frequency. The next two rows correspond to engine defects and exhibit anomalous peaks in their Fourier spectra.}
\end{figure}

Methods that utilize the approach based on the current signature analysis use a fast Fourier transform (FFT) to transfer data from the time to the frequency domain \cite{RomeroTroncoso2017}. The Fourier transform \( F(\omega) \) of a continuous time-domain signal \( x(t) \) can be expressed as the following integral:
\begin{equation}
F(\omega) = \int_{-\infty}^{\infty} x(t) e^{-i \omega t} \, dt,
\end{equation} 
where \( x(t) \) is the original signal as a function of time and \( \omega \) is the frequency.

Despite the aforementioned advantages, such as an unsupervised nature, the method has some limitations. For example, certain failures produce frequencies that are very close to the engine’s main operational frequency. As a result, these failures detection becomes challenging and poorly distinguishable. Furthermore, FFT-based methods struggle to distinguish harmonics caused by motor malfunctions from those caused by changes in speed, voltage, or load. In particular, under high load, certain types of faults are more clearly pronounced, as confirmed by the findings in \cite{Sharma2018}, which require more complicated methods such as windowed Fourier or Hilbert transforms \cite{hilbert}.

At the same time, discrete wavelet transform is widely used for the detection of rotor bar faults in induction motors due to its efficient multi-resolution analysis capabilities \cite{Ameid2018}. While the signature approach can be used for online analysis through modifications \cite{Jung2006}, it cannot incorporate multi-domain data sources, such as temperature, load, or other engine features, to enhance anomaly detection accuracy. On the other hand, machine learning techniques \cite{Choudhary2019} are expected to improve efficiency of online analysis as the number of information sources increases

\subsection{ML Approaches}
There are numerous articles that cover the application of artificial neural networks in the fault diagnosis of three-phase motors, ranging from vibration signal analysis \cite{Ribeiro2020} to motor current signature analysis for detecting bearing faults \cite{Dhomad2020}. But to date, the application of machine learning techniques in real-world induction motor fault detection remains underexplored due to the challenges discussed in the next section, with existing publications often relying on older methods like SVM \cite{SVM_art}. In contrast, this work is aimed for more recent solution studies like VAE-LSTM \cite{VAE-LSTM} or AnomlyBERT \cite{jeong2023anomalybert}, along with our own unsupervised approach.

One of the classical methods in the field of induction motor fault detection is the use of Support Vector Machines (SVM) \cite{SVM_art}. SVM operates by finding the hyperplane that best separates the different classes in the feature space with the maximum margin. The method allows to combine several features extracted from CSA and perform multiclass classification to distinguish between different fault types and does not require a lot of computational power, while providing high accuracy given enough high quality labeled data. For example, in \cite{SVM} authors use FFT to get frequency spectra and extract features and then use SVM to distinguish between five different classes. The SVM approach is then compared with a neural network. While the test accuracy of both models is pretty high, the SVM requires much lower training time.

Convolutional Neural Networks (CNNs) are also an effective approach for temporal data classification. A CNN is usually a deep neural network that uses convolutional layers to extract features by applying learnable filters to the input data. While this method is mostly used for image classification, its feature extraction capabilities can be advantageous in applications to induction motor data. For example, in the study \cite{2020convolutional} , a convolutional neural network was utilized to identify faults in the rotor bars of the induction motors. This approach combined current signature analysis (CSA) with analysis based on Fourier transform and demonstrated an outstanding classification accuracy across four different defect types.

\subsection{Existing ML approaches limitations}
To this point, from all the methods described above, it might seem that one could already effectively solve the problem of anomaly detection in three-phase induction motors. However, all of the existing solutions require a lot of labeled data consisting of both normal and defected observations. While collecting data from a healthy motor is a relatively  simple task\footnote{We already have an operational induction motor and can easily collect data.}, acquiring data corresponding to different types of faults would require conducting a costly intentional damage of the motor, or taking care of the collection of data from naturally occurring damages much in advance while there are still no good publicly available datasets in this field. Also, these supervised approaches usually tend to overfit on a specific engines and its utilization conditions, which may lead to a drastic performance degradation further.

The limitations of the existing machine learning approaches, as discussed earlier, highlight the necessity of adopting modern unsupervised methods that do not require anomalous observations in the training set.However, the outcomes of implementing and testing the VAE-LSTM \cite{VAE-LSTM} and AnomalyBERT \cite{jeong2023anomalybert} models, as presented in the further results section, underscore the limitations of these modern approaches, which hinder their applicability to the problem constraints outlined in this paper. Namely, the VAE-LSTM model still requires labeled observations to optimize the classification threshold and provides low accuracy due to the complex nature of anomalies. Meanwhile, AnomalyBERT requires no labeled data as it proposes a data degradation scheme that generates synthetic anomalies used to train the model. However, these anomalies are not related to the complex physical nature of the real data, preventing the trained model from effective detection during the testing stage.

In contrast to the aforementioned methods, this paper introduces a novel signature-guided unsupervised data augmentation technique aimed at enriching datasets with physically accurate anomalies. This approach addresses the issue of scarcity and unavailability of anomalous observations, enabling their use in latest supervised machine learning approaches.

\section{Methodology}

The proposed approach of signature-guided data augmentation (SGDA) draws inspiration from AnomalyBert \cite{jeong2023anomalybert}, employing a data degradation strategy that alters the input data to produce artificially labeled anomalies. As previously discussed, the main limitation of the AnomalyBert data degradation scheme lies in the simplicity and naivety of surrogate anomalies, as the proposed synthetic anomalies do not have physical interpretation behind. To address this issue and generate anomalies that are more meaningful in the context of induction motor time series, this study employs a novel anomaly sampling technique grounded on signature analysis approach. As discussed earlier, the method utilizes the idea that different types of engine faults result to specific current signatures, which have certain harmonics in the stator's spectrum. In other words, the main idea behind the signature analysis suggests that peaks on certain frequencies correspond to certain types of anomalies, and these frequencies can be pre-calculated on the basis of the physical characteristics of the induction motor. 

This means that if we use data collected from the engine to insert a peak at a frequency associated with a certain fault, we can produce a physically accurate synthetic observation that corresponds to a specific type of fault. This approach allows to augment the already existing dataset with new labeled observations, or create a new synthetic dataset without the need to damage the existing motor. What is more, we are able to apply supervised learning models, which was not possible before due to the absence of sufficient number of labeled data samples.

The SGDA approach introduced in this work can be divided into several steps. First, an FFT is applied to the segments of the data, which transforms the data from time to frequency domain. Next, the first 250 data points are extracted from each spectrum as they contain all necessary diagnostic frequencies. Then, the signature approach is applied to calculate the frequencies corresponding to specific faults based on the characteristics of the engine. In this work, we focus on five different types of faults including: rotor bar defect, air gap eccentricity, inter-turn short circuits, bearing defects of rolling element, bearing defects of outer race, bearing defects of inner race and other mechanical defects. However, the findings can be generalized to any number of faulty classes. After the frequencies have been calculated, the anomalous peaks are being injected at these frequencies into the training spectrum sample, while the resulting sample is labeled according to a type of injected defect. As a result of the procedure, a labeled dataset with surrogate defects is generated.

\subsection{Proposed method}
At the first stage, as discussed earlier, a signature analysis is utilized to estimate frequencies corresponding to specific anomalies given the physical parameters of the motor. Next, the segments are being extracted from the FFT-transformed time-series data. The segment length is set to 20 data points, and the target frequency of the augmentation peak is defined in the first step using a signature approach. All the extracted segments are then min-max normalized to be able to generate peaks in the same range compared to the original data.

A Variational Autoencoder (VAE) \cite{kingma2022autoencodingvariationalbayes} acts here as a generative model for learning synthetic peak segments that closely resemble the original segments. After the pre-training, the VAE is being used for synthetic peak segment generation. The generated peak segments are denormalized to the amplitude range of the original data by minimum and maximum values of the original segments. These denormalized synthetic peak segments are injected to the normal time series data at the target frequency area. For each data file to be processed, the generated peaks are inserted within the segment centered at the target frequency with a fixed segment length of 20 data points. The insertion is performed in such a way that the generated peak blends with the original data as if the peak was part of the normal recording.

During the experiments, a less complex generator-free augmentation approach is also studied. The approach is based on convolution with Gaussian kernel to smooth the simulated peaks.

For the final classification step, a model with ResNet \cite{he2015deepresiduallearningimage} architecture designed for time-series classification is used. Each block contains two 1D convolutional layers, where the number of filters in the layers is increased by a factor of two progressively, starting from 64 filters at the head to 512 at the tail. The model is trained with a batch size of 16, using the Adam optimizer with a learning rate of 0.001 and CrossEntropyLoss as the loss function.

Algorithm \ref{alg:Algo} provides a brief summary of the SGDA approach.

\begin{algorithm}[H]
\caption{Summary of SGDA Approach}
\label{alg:Algo}
\begin{algorithmic}[noend]

\State \textbf{Begin}

\State

\State \textbf{1. Data Collection}
    \State RawData $\leftarrow$ CollectDataFromMotor()

\State

\State \textbf{2. Estimate Anomalous Frequencies}
    \State AnomalousFrequencies $\leftarrow$ SignatureAnalysis(MotorParameters)

\State

\State \textbf{3. Data Preparation}
    \State Segments $\leftarrow$ SplitData(RawData)
    \For{each segment in Segments}
        \State FFTSegment $\leftarrow$ ApplyFFT(segment)
        \State FFTSegments.Append(FFTSegment)
    \EndFor

\State

\State \textbf{4. Segment Extraction}
    \For{each FFTSegment in FFTSegments}
        \State TargetSegment $\leftarrow$ ExtractSegment(FFTSegment, AnomalousFrequencies)
        \State NormalizedSegment $\leftarrow$ Normalize(TargetSegment)
        \State NormalizedSegments.Append(NormalizedSegment)
    \EndFor

\State

\State \textbf{5. Synthetic Anomaly Generation}
    \State \textbf{Option A:}
        \State VAEModel $\leftarrow$ TrainVAE(NormalizedSegments)
        \State GeneratedPeaks $\leftarrow$ GenerateSyntheticPeaks(VAEModel)
        \State SyntaticPeaks $\leftarrow$ Denormalize(GeneratedPeaks)
    \State \textbf{Option B:}
        \State SyntheticPeaks $\leftarrow$ GeneratePeaksWithConvolution(FFTSegments, GaussianKernelParameters)

\State

\State \textbf{6. Insertion of Synthetic Anomalies}
    \For{index $i$ in range(len(FFTSegments))}
        \State ModifiedSegment $\leftarrow$ InsertPeak(FFTSegments[$i$], SyntaticPeaks, AnomalousFrequencies)
        \State ModifiedSegments.Append(ModifiedSegment)
    \EndFor

\State

\State \textbf{7. Model Training}
    
    \State TrainedModel $\leftarrow$ TrainModel(ResNetModel, ModifiedSegments, Labels, TrainingParameters)

\State

\State \textbf{8. Anomaly Detection and Classification}
    \State Anomalies $\leftarrow$ DetectAnomalies(TrainedModel, NewData)

\State

\State \textbf{End}

\end{algorithmic}
\end{algorithm}

\section*{3 Results}
\section*{3.1 Data}
The experiments are conducted using three datasets collected from real setups. The first dataset contains only normal data collected from a healthy induction motor. The second dataset includes five different labeled anomalous files, which correspond to defected motor, and three files with healthy motor data. These normal and anomalous files in this dataset are produced by the same engine as in the first dataset, with known characteristics. The third dataset is generated by a different induction motor with unknown parameters and contains healthy data and anomalous data of unknown type.

\section*{3.2 Experimental Setups}
We test our SGDA approach under different conditions, evaluating the synthetic data generation produced by convolution with a Gaussian kernel and deep convolutional generative neural network. 

In the first experiment (Table \ref{tab:results_E1}), designed for binary classification on synthetic data, we injected anomalous peaks into the normal data at specific frequencies corresponding to the type of fault using the previously introduced technique. As a result, the testing dataset contains an equal proportion of real normal data and synthetic anomalous data. Since there is no limit on the number of normal observations, the healthy and synthetic anomalous classes are balanced in the training set. Thus, the model is trained and tested on a mixture of normal samples and synthetic anomalies.

In the second experiment (Table \ref{tab:results_E2}), we test five-class classification using synthetic anomalous data. Due to the limited availability of real anomalous data, we evaluate the SGDA approach in a multiclass classification setting using synthetic anomalies. These synthetic anomalous peaks are generated using a Gaussian kernel convolution to simulate the varying shapes of peaks associated with different types of faults. For each type of fault, peaks are injected into the normal data at frequencies determined by the signature approach. The classes are balanced in the training set for all models, and along with SGDA-generated data.

In the third experiment (Table \ref{tab:results_E3}), we perform a binary classification task on real data, attempting to distinguish normal data from anomalous data with an inter-turn short circuit. In the fourth experiment (Table \ref{tab:results_E4}), we conduct a three-class classification task on real data, where, in addition to detecting the inter-turn short circuit, the model has to identify anomalies corresponding to mechanical defects, such as rotor imbalance. In both experiments, we train a generative neural network on peaks extracted from the files corresponding to the inter-turn short circuit of phase A or the file corresponding to rotor imbalance. The training set consists of normal data from the second dataset, small portion of real data, and synthetically generated data. The model is tested using normal data, real anomalous data, and a portion of synthetic data too. Class balance is maintained in both the training and testing datasets.

In all the experiments, the ResNet-based classification model is used. Data is preprocessed with an FFT-transformed rolling window.

The SGDA model is compared with AnomalyBERT \cite{jeong2023anomalybert}, VAE-LSTM \cite{VAE-LSTM}, MLP~\cite{MLP},  and SVM~\cite{SVM}.

\section*{3.3 Results}

In this section, the results for four different types of experiments are presented. Two experiments are based solely on synthetic data are included to evaluate the approaches in the absence of real anomalous data, focusing on both binary and multiclass classification (see Table \ref{tab:results_E1} and Table \ref{tab:results_E2}, respectively). The models are also tested on real anomalous (defect) data for binary classification (Table \ref{tab:results_E3}) and multiclass classification (Table \ref{tab:results_E4}). The tables below present the results of the proposed SGDA approach, two unsupervised baseline models (AnomalyBERT and VAE-LSTM) and the supervised SVM baseline.

\begin{table}[H]
\centering
\begin{tabular}{lcc}
\toprule
\textbf{Method} & \textbf{Accuracy (\%)} & \textbf{F1-score} \\
\midrule
SGDA & $100.00\pm0.01$ & $1.00\pm0.01$ \\
AnomalyBERT & $54.43\pm2.16$ & $0.53\pm0.03$ \\
VAE-LSTM & $60.49\pm1.22$ & $0.59\pm0.02$ \\
SVM & $57.41\pm3.05$ & $0.56\pm0.02$ \\
\bottomrule
\end{tabular}
\caption{Binary classification with synthetic data.}
\label{tab:results_E1}
\end{table}

In Table \ref{tab:results_E1}, the results for the basic binary classification on synthetic data are presented.Anomalous data for testing, as part of the SGDA approach, was generated based on the signature method. We can see that the proposed SGDA approach achieves the best metrics, with an accuracy of $100$\% and an F1-score of $1.00$, while the baseline unsupervised approaches (AnomalyBERT, VAE-LSTM, and SVM) achieve much lower accuracies, ranging from $54.43$\% to $60.49$\%.

\begin{table}[H]
\centering
\begin{tabular}{lcc}
\toprule
\textbf{Method} & \textbf{Accuracy (\%)} & \textbf{F1-score} \\
\midrule
SGDA & $99.51\pm0.03$ & $0.99\pm0.03$ \\
AnomalyBERT & -- & -- \\
VAE-LSTM & -- & -- \\
SVM & $21.16\pm2.07$ & $0.06\pm0.01$ \\
\bottomrule
\end{tabular}
\caption{Five-class classification with syntactic data. AnomalyBERT and VAE-LSTM scores are empty due to their inapplicability to the multiclass synthetic data.}
\label{tab:results_E2}
\end{table}

In Table \ref{tab:results_E2}, the results for the five-class classification task are presented. In this experiment, anomalous observations were generated by integrating peaks at frequencies specified by the signature approach, corresponding to four different types of anomalies. Here, no anomalous samples are being provided to the models during the training process. As part of the SGDA approach, anomalous data is generated based on the signature method to simulate the anomalies during training. One can see that SGDA maintains a high accuracy of $99.51$\% and an F1-score of $0.99$, demonstrating strong performance even in the absence of "real" anomalous data during training. In contrast, the SVM method shows very low accuracy at $21.16$\% and an F1-score of $0.06$, as it struggles due to the lack of access to anomalous data. The scores for AnomalyBERT and VAE-LSTM are not provided, as these unsupervised approaches cannot be straightforwardly applied to this multiclass synthetic data task.

\begin{table}[H]
\centering
\begin{tabular}{lcc}
\toprule
\textbf{Method} & \textbf{Accuracy (\%)} & \textbf{F1-score} \\
\midrule
SGDA & $99.45\pm0.13$ & $0.99\pm0.02$ \\
AnomalyBERT & $53.26\pm1.23$ & $0.52\pm0.05$ \\
VAE-LSTM & $59.12\pm2.51$ & $0.58\pm0.07$ \\
SVM & $55.13\pm2.32$ & $0.54\pm0.01$ \\
\bottomrule
\end{tabular}
\caption{Binary classification with real anomalous data.}
\label{tab:results_E3}
\end{table}

Table \ref{tab:results_E3} presents the results for binary classification of real anomalous data. In this experiment, during the testing phase, the models are aimed to distinguishing between real normal and anomalous data. However, real anomalous data is not available to any of the models during training. Instead, synthetic anomalous data is generated as a part of the SGDA approach to train the model. The SGDA method outperforms the others, achieving an accuracy of $99.45$\% and an F1-score of $0.99$. In contrast, AnomalyBERT, VAE-LSTM, and SVM show significantly lower performance, with accuracies ranging from $53.26$\% to $59.12$\% and F1-scores between $0.52$ and $0.58$.

\begin{table}[H]
\centering
\begin{tabular}{lcc}
\toprule
\textbf{Method} & \textbf{Accuracy (\%)} & \textbf{F1-score} \\
\midrule
SGDA & $98.75\pm0.14$ & $0.98\pm0.02$ \\
AnomalyBERT & -- & -- \\
VAE-LSTM & -- & -- \\
SVM & $34.52\pm2.84$ & $0.16\pm0.01$ \\
\bottomrule
\end{tabular}
\caption{Three-class classification with real anomalous data. AnomalyBERT and VAE-LSTM scores are empty due to their inapplicability to the multiclass setting.}
\label{tab:results_E4}
\end{table}

Table \ref{tab:results_E4} presents the results of the experiment where the models are used to distinguish two different types of real anomalies and normal observations. The real anomalies used to test the models are not accessible during training. For the SGDA approach, anomalous data is generated based on the signature approach corresponding to the types of faults present in the testing set. As shown, the SGDA method achieves a high accuracy of $98.75$\% and an F1-score of $0.98$, demonstrating its effectiveness in handling the multiclass classification task even without access to real anomalies during training. In contrast, the SVM model achieves significantly lower performance, with an accuracy of $34.52$\% and an F1-score of $0.16$, indicating its inability to generalize well under these conditions. AnomalyBERT and VAE-LSTM scores are not reported due to their inapplicability to the multiclass setting.

\section*{3.4 Epsilon Testing}

We also investigate the influence of the number of available anomalous observations in the training set on the quality of the model. Specifically, we compare the accuracy of the proposed approach of signature-guided data augmentation (SGDA) \footnote{Which requires no real anomalous observations in the training set.} with that of some of the most popular approaches to anomaly detection in induction motors.

The first model we consider is a convolutional neural network (CNN) that has been previously used in the area of anomaly detection in time series, for example by \cite{2020convolutional}. We also compare the proposed approach with support vector machines (SVMs) \cite{VAE-LSTM}, which are often used to detect anomalies in the field of induction motors \cite{sunal2022review}. The last model is a multi-layer perceptron (MLP) inspired by the architecture used in \cite{MLP} to detect failures associated with bearing faults.

Table \ref{tab:results_epsilon} presents a comparison of the accuracy of the chosen models in the task of binary classification, depending on the number of available anomalous observations in the training set.

\begin{table}[h!]
    \centering
    \resizebox{\textwidth}{!}{%
    \begin{tabular}{lcccccc}
        \toprule
        \# of Observations & 1 & 10 & 25 & 50 & 100 & 300 \\
        \midrule
        CNN & $50.00\pm0.00$ & $66.50\pm17.40$ & $90.90\pm13.95$ & $97.19\pm7.15$ & $98.69\pm3.39$ & $99.62\pm1.51$ \\
        SVM & $62.27\pm2.96$ & $97.09\pm2.89$ & $98.82\pm1.31$ & $99.72\pm0.28$ & $99.87\pm0.12$ & $99.94\pm0.09$ \\
        MLP & $50.00\pm0.00$ & $50.23\pm0.70$ & $50.47\pm0.81$ & $51.42\pm1.41$ & $51.67\pm1.44$ & $52.71\pm1.86$ \\
        \textbf{SGDA} & $\bm{99.98\pm0.04}$ & $\bm{99.98\pm0.04}$ & $\bm{99.98\pm0.04}$ & $\bm{99.98\pm0.04}$ & $\bm{99.98\pm0.04}$ & $\bm{99.98\pm0.04}$ \\
        \bottomrule
    \end{tabular}%
    }
    \caption{Comparison of Model Accuracy (\%) Based on the Number of Anomalous Observations. This table shows the performance of four different models when varying the number of anomalous observations used in the training dataset. The accuracy is presented with the mean and standard deviation for each model across different observation counts.}
    \label{tab:results_epsilon}
\end{table}

As shown in Table \ref{tab:results_epsilon}, the proposed approach consistently achieves high accuracy regardless of the number of anomalous observations included in the training set, maintaining an accuracy of \textbf{$99.98\% \pm 0.04\%$} across all the scenarios. 

The CNN model's accuracy rises from \textbf{$50.00\% \pm 0.00\%$} with one anomalous observation to \textbf{$99.62\% \pm 1.51\%$} with 300 observations. Similarly, the SVM model's accuracy improves from \textbf{$62.27\% \pm 2.96\%$} to \textbf{$99.94\% \pm 0.09\%$} as more anomalous data being included.

The MLP model, however, demonstrates the same level of accuracy regardless of the number of anomalous observations.

\clearpage
\section*{4 Discussion}

From the four tables presented in section 3.3, we can see the proposed algorithm demonstrates high performance across wide range of scenarios, including noisy training datasets with limited or absent anomalous observations. The proposed approach achieves high accuracy when tested on real data too, supporting its potential applications in real-world cases. In contrast, adapted state-of-the-art unsupervised approaches demonstrate limited applicability and low accuracy due to the high level of noise in the data and the complex nature of real anomalies.

Results of epsilon testing demonstrate the proposed method does not significantly rely on real anomalous observations during training, making it highly effective even when such data is scarce or unavailable. It also supports the newly sampled anomalies are consistent with real ones. In contrast, the CNN and SVM models show significant improvements in accuracy as the number of anomalous observations increases. The accuracy of the MLP model does not change as more anomalous observations are introduced. This is because adapting the MLP model effectively requires specific data preprocessing and feature generation tailored to the anomalies. Without the ability to adjust the approach for data preprocessing and variable generation\footnote{Due to the lack of anomalous observations in our experiment}, the MLP cannot learn the distinguishing patterns necessary for accurate classification. In our context, it is impossible to perform such adaptations, as we assume the absence of anomalous data during training. This limitation highlights the challenges traditional models face when anomalous data is unavailable and emphasizes the necessity of methods like ours that do not depend on such data. In contrast, the introduced algorithm demonstrates consistently high outperformance on this experiment too.

The proposed approach solves several problems that exist in the area of anomaly detection in induction motors. First, it allows the generation of labeled anomalous data based on normal data recorded from the engine, without the need to damage the induction motor in order to collect data corresponding to different types of faults. Moreover, it provides better accuracy and more informative predictions than the existing unsupervised solutions. It also allows the implementation of various supervised machine learning models and enables training them on synthetic anomalies, ensuring the model perform well with real-world anomalies. At the same time, the proposed approach is much cheaper and easier to implement than other published supervised learning approaches, as it requires no special experimental setup to generate anomalous data. The data can be recorded from an already installed and working induction motor by simply attaching current clamps. Finally, the proposed approach extends the signature approach to multidimensional datasets by allowing users to add various sources of data during training and testing stages, thus improving the robustness and accuracy of the predictions.

While the experimental results suggest high accuracy of the proposed approach in real-world scenarios, in some extreme cases, synthetic anomalies may not fully capture the complexity of real-world faults, which could slightly reduce its effectiveness when applied to certain data. Additionally, the approach depends heavily on accurate motor parameters to calculate fault-related frequencies. In cases, where these parameters are unavailable or difficult to determine, the accuracy of the synthetic anomalies (and thus the overall effectiveness of the model) may be compromised. To tackle this, we believe the more sophisticated semi-supervised generative techniques for the anomalies sampling can be used, increasing the generated anomalies quality and variety in these scenarios too.

\section{Conclusion}
In this research, we adapt several unsupervised machine learning approaches for the problem of anomaly detection in induction motors. We show that  due to the limitations of these models in a case of limited labeled samples number, the results remain suboptimal. To mitigate the issue, we develop new physically-guided unsupervised approach that combines signature analysis, a generative neural networks power, and a ResNet-based classification model. The results show that the proposed algorithm consistently outperforms other methods in detecting anomalies in induction motors, achieving high accuracy even in challenging scenarios with noisy data and few anomalous observations. With all the aforementioned points, the proposed approach demonstrates high potential for further research along with a wide application range to real motor-diagnostic challenges.

\clearpage
\section*{Data Availability Statement}
The datasets analyzed during the current study are available from the corresponding author upon reasonable request. Please note that, due to confidentiality agreements with the external company that provided the data, the datasets cannot be made publicly available.
\section*{Conflicts of Interest}
This publication was supported by the grant for research centers in the field of AI provided by the Analytical Center for the Government of the Russian Federation (ACRF) in accordance with the agreement on the provision of subsidies (identifier of the agreement 000000D730321P5Q0002) and the agreement with HSE University No. 70-2021-00139. The funders had no role in study design, data collection and analysis, decision to publish, or preparation of the manuscript.

Apart from this, the authors have no conflicts of interest to declare that are relevant to the content of this article.
 
\bibliographystyle{elsarticle-num}
\bibliography{ref}

\clearpage 
\appendix
\section{Leaky ReLU Activation Function and KL Divergence}

\subsection{Leaky ReLU Activation Function}

The Leaky Rectified Linear Unit (Leaky ReLU) addresses the "dying ReLU" problem by allowing a small, non-zero gradient when the input is negative. It is defined as:

\begin{equation}
f(x) = \begin{cases}
x, & \text{if } x \geq 0, \\
\alpha x, & \text{if } x < 0,
\end{cases}
\end{equation}

where $\alpha$ is a small constant, typically $\alpha = 0.01$.

\subsection{Kullback-Leibler Divergence}

The Kullback-Leibler (KL) divergence measures how one probability distribution $P$ diverges from a reference distribution $Q$. For discrete distributions:

\begin{equation}
D_{\text{KL}}(P \parallel Q) = \sum_{i} P(i) \log \left( \frac{P(i)}{Q(i)} \right).
\end{equation}

For continuous distributions:

\begin{equation}
D_{\text{KL}}(P \parallel Q) = \int_{-\infty}^{\infty} P(x) \log \left( \frac{P(x)}{Q(x)} \right) dx.
\end{equation}

\end{document}